# An Equivalent ABCD-Matrix Formalism for Non-Local Wire Media with Arbitrary Terminations

Alexander B. Yakovlev, *Senior Member, IEEE*, Mário G. Silveirinha, *Fellow, IEEE*, George W. Hanson, *Fellow, IEEE*, and Chandra S. R. Kaipa

*Abstract*— A simple analytical model based on the transmission-matrix approach is proposed for equivalent wire-medium (WM) interfaces. The obtained ABCD matrices for equivalent interfaces capture the non-local effects due to the evanescent transverse magnetic (TM) WM mode and in part due to the propagating transverse electromagnetic (TEM) WM mode. This enables one to characterize the overall response of bounded WM structures by cascading the ABCD matrices of equivalent WM interfaces and WM slabs as transmission lines supporting only the propagating TEM WM mode, resulting in a simple circuit-model formalism for bounded WM structures with arbitrary terminations, including the open-end, patch/slot arrays, and thin metal/2D material, among others. The individual ABCD matrices for equivalent WM interfaces apparently violate the conservation of energy and reciprocity, and therefore, the equivalent interfaces apparently behave as non-reciprocal lossy or active systems. However, the overall response of a bounded WM structure is consistent with the lossless property maintaining the conservation of energy and reciprocity. These unusual features are explained by the fact that in the non-local WM the Poynting vector has an additional correction term which takes into account a "hidden power" due to non-local effects. Results are obtained for various numerical examples demonstrating a rapid and efficient solution for bounded WM structures, including the case of geometrically complex multilayer configurations with arbitrary terminations, subject to the condition that WM interfaces are decoupled by the evanescent TM WM mode below the plasma frequency.

## I. Introduction

At microwave frequencies wire media as an artificial material has been known for a long time [1], [2], and it has gained attention in the last two decades in metamaterials research ranging from microwaves to optics in relation to observed anomalous wave phenomena such as negative refraction [3]-[9], canalization, transport, and magnification of the near field to distances of several wavelengths [10]-[14], subwavelength imaging of the near field [15]-[22], and radiative heat transfer [23]-[25]. A broad range of applications of wire-media metamaterials at terahertz (THz) and optical frequencies is given in [26].

Manuscript received April XX, 2019. This work was partially supported by (M.S.) by Fundação para Ciência e a Tecnologia (FCT) under project UID/EEA/50008/2019, and by the European Regional Development Fund (FEDER), through the Competitiveness and Internationalization Operational Programme (COMPETE 2020) of the Portugal 2020 framework, Project, RETIOT, POCI-01-0145-FEDER-016432.

A. B. Yakovlev is with the Department of Electrical Engineering, University of Mississippi, University, MS 38677-1848, USA. (email: yakovlev@olemiss.edu).
M. G. Silveirinha is with the Instituto Superior Técnico, University of Lisbon and Instituto de Telecomunicações, Torre Norte, Av. Rovisco Pais 1, Lisbon 1049-001, Portugal. (email: mario.silveirinha@co.it.pt).
G. W. Hanson is with the Department of Electrical Engineering and Computer Science, University of Wisconsin-Milwaukee, Milwaukee, Wisconsin 53211, USA. (email: george@uwm.edu).
C. S. R. Kaipa is with Qualcomm Technologies Inc., 5745 Pacific Center Blvd., San Diego, CA 92121, USA. (email: cskaipa@gmail.com).

In addition, WM metamaterials have been utilized in various applications at microwave frequencies, including high-impedance mushroom-type substrates as electromagnetic band-gap surfaces for low-profile antennas [27]-[31], broadband high-impedance surface absorbers with stable angle characteristics [32]-[34], epsilon-near-zero metamaterials [35]-[37], and gap waveguide technology [38]-[43], among others.

It is already well known that at microwave frequencies, even in the very long wavelength limit, wire media is characterized by strong spatial dispersion (SD) effects [44], [45], such that the constitutive relations between the macroscopic fields and the electric dipole moment are non-local. The role of SD in the analysis of electromagnetic interaction with wire media has been addressed, resulting in the development of non-local homogenization formalism [29], [30], [46], [47] which necessitates the use of additional boundary conditions (ABCs) at WM terminations [34], [48]-[52]. Various non-local homogenization methods have been developed for excitation, radiation, and scattering electromagnetic problems involving wire media and WM-type structures [29], [34], [47], [53]-[61]. In the above publications the importance of non-local homogenization for WM-type structures has been established, unless the SD effects are suppressed or significantly reduced as in the mushroom topologies (with electrically short wires) where the local model formalism can provide physical results [29], [30], [62].

The analysis proposed here concerns the development of an equivalent transmission (ABCD) matrix approach for non-local WM interfaces. This work was triggered by recent advancements in modeling of non-local material interfaces, wherein the smearing of surface charges due to non-local effects was approximated by a local, finite-thickness layer [63], [64]. In [63] it was shown that the spatial non-locality in metals can be represented by a composite material comprising a thin local dielectric layer on top of a local metal. Moreover, in [64] a local thickness-dependent permittivity was derived in closed form for bounded non-local WM structures, which takes into account the SD effects as an average per length of the wires and the effect of the boundary. In this regard we should point out Ref. [65] where metamaterial effective parameters that depend on geometry have been discussed. In [63]-[65] the non-local effects are captured in a subwavelength effective dielectric layer. However, in general in WM the SD effects are not confined at the interface; they are distributed through the entire non-local material due to presence of two extraordinary waves: the TM mode, which is evanescent below the plasma frequency of the WM, and the TEM mode, which propagates in WM as in an uniaxial material with *extreme anisotropy*.

We propose a simple analytical model based on the ABCD-matrix approach for equivalent WM interfaces to capture the non-local effects due to the evanescent TM WM mode and in part the non-local effects due to the propagating TEM WM mode, with the rest of the material supporting only the propagating TEM WM mode. Two semi-infinite structures are considered with equivalent WM interfaces: (i) semi-infinite local dielectric – non-local WM and (ii) two semi-infinite non-local WM. In both cases the WM in general can be terminated with an impedance surface (open-end, patch/slot arrays, thin metal/2D material, among others). It is observed that in such approach with the equivalent interfaces (local – non-local materials and two different non-local WM) leads to a formalism that is apparently non-reciprocal due to the unusual form of the Poynting vector in nonlocal media. This is a critical point in developing an equivalent interface for a WM used in modeling bounded WM structures with impedance-surface terminations and in the multilayered WM environment. The ABCD matrices for equivalent WM interfaces are retrieved from the conventional and additional boundary conditions depending on the WM termination. The response of an entire bounded WM structure due to an obliquely incident TM-polarized plane wave is modeled by cascading the ABCD matrices at the equivalent WM interfaces and the WM slabs as transmission lines supporting the only TEM WM propagating mode. It is noted that the ABCD matrix for an equivalent WM interface seemingly violates the conservation of energy and reciprocity, and therefore, apparently the interface behaves as a non-reciprocal lossy or active system. The key point is that there are always at least two interfaces in a bounded WM, and so if one of them provides loss (as the wave enters the WM), the other interface provides gain (as the wave exits the WM), such that the overall response is consistent with the lossless property maintaining the conservation of energy and reciprocity. The apparent "gain" and "loss" are explained by the fact that in the non-local WM the Poynting vector has an additional correction term corresponding to the "hidden power" due to non-local effects in the WM [57]. We should also point out that work which is closely related to the material of this paper has been presented in [65], however, in [65] the equivalent network analysis is carried out for a three-port network, wherein the transmission lines corresponding to the modes in the local dielectric material and the non-local WM are coupled at the interface by the ABC, with the aim of deriving the ABCD matrix for a non-local WM (supporting TM and TEM extraordinary modes). This is different from the analysis of the current paper with the goal of deriving the ABCD matrices for equivalent WM interfaces.

Results are obtained for various WM configurations demonstrating excellent agreement with the non-local solution subject to the condition that the WM interfaces are decoupled by the evanescent TM WM mode below the plasma frequency. The proposed formalism of an equivalent transmission matrix is generalized for a multilayered WM with arbitrary impedance-surface terminations at the WM interfaces, simplifying the analysis of geometrically complex WM structures.

The paper is organized as follows. In Section II the equivalent transmission-network analysis is presented for WM interfaces with the discussion of the additional correction term in the Poynting vector in WM. In Section III various numerical examples of single-layer and multi-layer WM structures with different impedance-surface terminations at WM interfaces are presented based on the ABCD-matrix formalism and compared with the non-local solution. The conclusions are drawn in Section IV. Also, the paper is accompanied by two appendices with the analytical details concerning the additional correction term in the Poynting vector of WM, and derivation of the ABCD matrix for an equivalent interface of two non-local WM connected by an impedance surface. A time dependence of the form $e^{j\omega t}$ is assumed and suppressed.

II. ABCD-MATRIX APPROACH FOR EQUIVALENT WM INTERFACES

A. *Semi-Infinite Local Dielectric – Non-Local WM Equivalent Interface*

Consider a semi-infinite uniaxial non-local WM ($z < 0$) terminated with an impedance surface at an interface with a semi-infinite local dielectric material ($z > 0$) as shown in Fig. 1. The host permittivity of WM is $\varepsilon_h$, the permittivity of dielectric material is $\varepsilon_d$, the period of vias in the 2D square lattice is $a$, and the radius of vias is $r_0$. The TM-polarized plane wave is characterized with $H_y, E_x, E_z$ field components, and it can be incident either from the side of the dielectric material as the usual TEM wave or from the side of the WM as the extraordinary TEM WM mode. Note that for long wavelengths the TM WM mode is evanescent (with an exponential decay), and thereby the interfaces of sufficiently thick wire media structures are mainly coupled through the TEM WM mode.

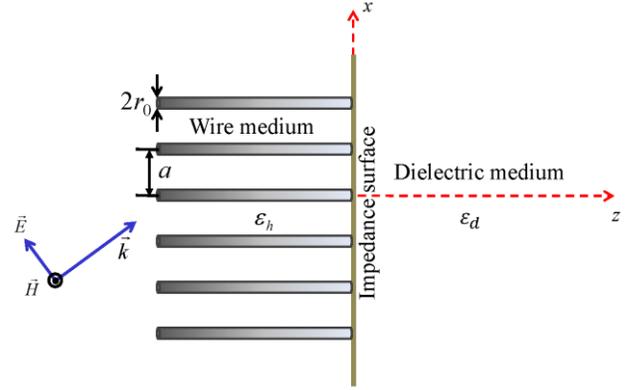

Fig. 1. Semi-infinite uniaxial non-local WM terminated with an impedance surface at an interface with a semi-infinite local dielectric material.

At the interface ($z = 0$) the fields satisfy the continuity of the tangential electric-field components, jump condition for the tangential magnetic-field components, and the generalized ABC [51], [52] for the surface current density $J_{z,\text{wm}}(z)$:

$$E_{x,\text{d}} = E_{x,\text{TEM}} + E_{x,\text{TM}} \tag{1a}$$

$$H_{y,\text{d}} = H_{y,\text{TEM}} + H_{y,\text{TM}} - Y_g (E_{x,\text{TEM}} + E_{x,\text{TM}}) \tag{1b}$$

$$J_{z,\text{wm}}(z) + \alpha \frac{dJ_{z,\text{wm}}(z)}{dz} = 0. \tag{1c}$$

Here, $Y_g$ is the surface admittance of an impedance surface with the closed-form expressions for printed and slotted inductive and capacitive grids given in [67]. The parameter $\alpha$ depends on the material properties of the impedance surface, and for metallic patches, $\alpha = C_p / C_w$, where $C_w$ and $C_p$ are given in [56], and for a thin metal/2D material characterized by the surface conductivity $\sigma_s$, $\alpha = \sigma_s / j\omega\varepsilon_0\varepsilon_h$ [52]. The ABC (1c) can be expressed in terms of the field components in the WM as follows:

$$k_x H_{y,\text{wm}} + \omega\varepsilon_0\varepsilon_h E_{z,\text{wm}} + \alpha \left( k_x \frac{\partial H_{y,\text{wm}}}{\partial z} + \omega\varepsilon_0\varepsilon_h \frac{\partial E_{z,\text{wm}}}{\partial z} \right) = 0. \tag{2}$$

Taking into account $E_{z,\text{wm}} \equiv E_{z,\text{TM}} = \frac{1}{j\omega\varepsilon_0\varepsilon_h\varepsilon_{zz}^{\text{TM}}} \frac{\partial H_{y,\text{TM}}}{\partial x}$, where $\varepsilon_{zz}^{\text{TM}} = 1 - k_p^2 / (k_p^2 + k_x^2)$, and $\frac{\partial}{\partial x} = -jk_x$, we obtain,

$$E_{z,\text{TM}} = -\frac{1}{\omega\varepsilon_0\varepsilon_h}\frac{k_p^2 + k_x^2}{k_x}H_{y,\text{TM}}. \quad (3)$$

Here, $k_p$ is the plasma wavenumber defined in [45, eq. (10)] and $k_x$ is the $x$-component of the wave vector $\mathbf{k} = (k_x, 0, k_z)$. Then, from Maxwell's equations, $\frac{\partial H_{y,\text{wm}}}{\partial z} = -j\omega\varepsilon_0\varepsilon_h\left(E_{x,\text{TEM}} + E_{x,\text{TM}}\right)$, and with the assumption that there is no incident TM mode on the WM interface and the reflected TM mode from the WM interface is in the negative $z$-direction (Fig. 1),

$$E_{x,\text{TM}} = -\frac{1}{j\omega\varepsilon_0\varepsilon_h}\frac{\partial H_{y,\text{TM}}}{\partial z} = \frac{j\gamma_{\text{TM}}}{\omega\varepsilon_0\varepsilon_h}H_{y,\text{TM}} \quad (4)$$

and that $\frac{\partial E_{z,\text{wm}}}{\partial z} \equiv \frac{\partial E_{z,\text{TM}}}{\partial z} = -\frac{\gamma_{\text{TM}}}{\omega\varepsilon_0\varepsilon_h}\frac{k_p^2 + k_x^2}{k_p^2}H_{y,\text{TM}}$, the ABC (2) can be written as follows:

$$k_x(H_{y,\text{TEM}} + H_{y,\text{TM}}) - \frac{k_p^2 + k_x^2}{k_x}H_{y,\text{TM}}$$
$$+\alpha\left(-j\omega\varepsilon_0\varepsilon_h k_x(E_{x,\text{TEM}} + E_{x,\text{TM}}) - \gamma_{\text{TM}}\frac{k_p^2 + k_x^2}{k_x}H_{y,\text{TM}}\right) = 0. \quad (5)$$

Substituting (4) in (5) and after simplifications we obtain,

$$H_{y,\text{TM}} = \frac{k_x^2 H_{y,\text{TEM}} - j\omega\varepsilon_0\varepsilon_h k_x^2 \alpha E_{x,\text{TEM}}}{k_p^2(1 + \alpha\gamma_{\text{TM}})} \quad (6)$$

where $\gamma_{\text{TM}} = \sqrt{k_p^2 + k_x^2 - k_h^2}$ is the propagation constant of the TM WM mode, $k_h = k_0\sqrt{\varepsilon_h}$ is the wavenumber of the host medium, and $k_0 = \omega/c$ is the wavenumber of free space. Next, using (4) and (6) in the continuity equation for the tangential electric-field components (1a) results in,

$$E_{x,d} = \left(1 + \frac{k_x^2}{k_p^2}\frac{\alpha\gamma_{\text{TM}}}{(1 + \alpha\gamma_{\text{TM}})}\right)E_{x,\text{TEM}} + \frac{j\eta_h\gamma_{\text{TM}}}{k_h}\frac{k_x^2}{k_p^2}\frac{1}{(1 + \alpha\gamma_{\text{TM}})}H_{y,\text{TEM}} \quad (7)$$

and with (4) and (6) substituted in the jump condition for the tangential magnetic-field components (1b) we obtain,

$$H_{y,d} = \left(-Y_g - j\frac{k_h}{\eta_h}\frac{k_x^2}{k_p^2}\frac{\alpha}{(1 + \alpha\gamma_{\text{TM}})}\left(1 - Y_g\frac{j\eta_h\gamma_{\text{TM}}}{k_h}\right)\right)E_{x,\text{TEM}}$$
$$+\left(1 + \frac{k_x^2}{k_p^2}\frac{1}{(1 + \alpha\gamma_{\text{TM}})}\left(1 - Y_g\frac{j\eta_h\gamma_{\text{TM}}}{k_h}\right)\right)H_{y,\text{TEM}} \quad (8)$$

where $\eta_h$ is the intrinsic impedance of the host medium.

With (7) and (8) we can write for the WM interface at $z = 0$ shown in Fig. 1,

$$\begin{pmatrix} E_{x,d} \\ H_{y,d} \end{pmatrix} = \mathbf{M}_1 \cdot \begin{pmatrix} E_{x,\text{TEM}} \\ H_{y,\text{TEM}} \end{pmatrix} \quad (9)$$

where $\mathbf{M}_1$ is the ABCD matrix for an equivalent WM interface,

$$\mathbf{M}_1 = \begin{pmatrix} 1 + \frac{k_x^2}{k_p^2}\frac{\alpha\gamma_{\text{TM}}}{(1+\alpha\gamma_{\text{TM}})} & \frac{j\eta_h\gamma_{\text{TM}}}{k_h}\frac{k_x^2}{k_p^2}\frac{1}{(1+\alpha\gamma_{\text{TM}})} \\ -Y_g - j\frac{k_h}{\eta_h}\frac{k_x^2}{k_p^2}\frac{\alpha}{(1+\alpha\gamma_{\text{TM}})}\left(1 - Y_g\frac{j\eta_h\gamma_{\text{TM}}}{k_h}\right) & 1 + \frac{k_x^2}{k_p^2}\frac{1}{(1+\alpha\gamma_{\text{TM}})}\left(1 - Y_g\frac{j\eta_h\gamma_{\text{TM}}}{k_h}\right) \end{pmatrix}. \quad (10)$$

The ABCD matrix $\mathbf{M}_1$ for an equivalent interface captures the non-local effects due to the evanescent TM WM mode and in part the non-local effects due to the propagating TEM WM mode, with the rest of a semi-infinite WM supporting the only propagating TEM WM mode. Note that the theory is exact within the assumption that there is no incident TM WM mode. Because of the evanescent nature of the TM WM mode this assumption is typically very good in realistic structures with a finite thickness.

The ABCD matrix $\mathbf{M}_1$ for a special case of an open-end WM interface is obtained from (10) with $Y_g = 0$ and $\alpha = 0$,

$$\mathbf{M}_1 = \begin{pmatrix} 1 & \frac{j\eta_h\gamma_{\text{TM}}}{k_h}\frac{k_x^2}{k_p^2} \\ 0 & \frac{k_p^2 + k_x^2}{k_p^2} \end{pmatrix}. \quad (11)$$

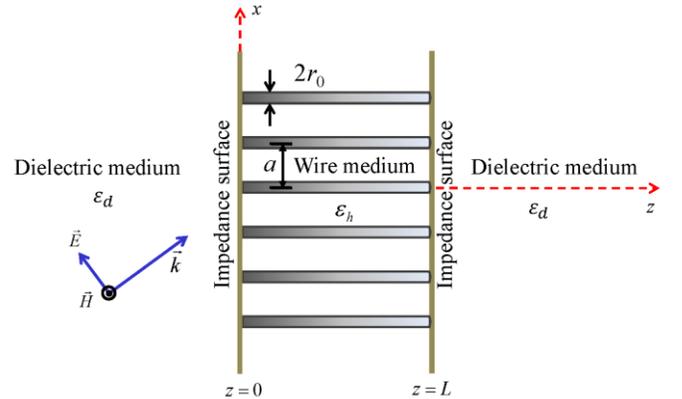

Fig. 2. Scattering of the TM-polarized plane wave from the WM slab terminated by impedance surfaces at both interfaces with the dielectric material.

For a finite WM slab terminated with impedance surfaces at both interfaces at $z = 0$ and $z = L$ (shown in Fig. 2) with the TM-polarized plane wave incident from the side of the local dielectric material ($z < 0$), the ABCD matrix $\mathbf{M}_1$ at the interface at $z = 0$ is obtained by changing $\alpha$ to $-\alpha$, $\gamma_{\text{TM}}$ to $-\gamma_{\text{TM}}$, and $Y_g$ to $-Y_g$ in (10),

$$\mathbf{M}_1 = \begin{pmatrix} 1 + \frac{k_x^2}{k_p^2}\frac{\alpha\gamma_{\text{TM}}}{(1+\alpha\gamma_{\text{TM}})} & -\frac{j\eta_h\gamma_{\text{TM}}}{k_h}\frac{k_x^2}{k_p^2}\frac{1}{(1+\alpha\gamma_{\text{TM}})} \\ Y_g + j\frac{k_h}{\eta_h}\frac{k_x^2}{k_p^2}\frac{\alpha}{(1+\alpha\gamma_{\text{TM}})}\left(1 - Y_g\frac{j\eta_h\gamma_{\text{TM}}}{k_h}\right) & 1 + \frac{k_x^2}{k_p^2}\frac{1}{(1+\alpha\gamma_{\text{TM}})}\left(1 - Y_g\frac{j\eta_h\gamma_{\text{TM}}}{k_h}\right) \end{pmatrix}. \quad (12)$$

The ABCD matrix $\mathbf{M}_2$ for an equivalent WM interface at $z = L$ can be written as the inverse matrix of $\mathbf{M}_1$ given by (10),

$$\mathbf{M}_2 = \begin{pmatrix} 1 + \dfrac{k_x^2}{k_p^2} \dfrac{\alpha \gamma_{\mathrm{TM}}}{(1+\alpha\gamma_{\mathrm{TM}})} & \dfrac{j\eta_h \gamma_{\mathrm{TM}}}{k_h} \dfrac{k_x^2}{k_p^2} \dfrac{1}{(1+\alpha\gamma_{\mathrm{TM}})} \\ -Y_g - j\dfrac{k_h}{\eta_h}\dfrac{k_x^2}{k_p^2}\dfrac{\alpha}{(1+\alpha\gamma_{\mathrm{TM}})}\left(1 - Y_g \dfrac{j\eta_h\gamma_{\mathrm{TM}}}{k_h}\right) & 1 + \dfrac{k_x^2}{k_p^2}\dfrac{1}{(1+\alpha\gamma_{\mathrm{TM}})}\left(1 - Y_g \dfrac{j\eta_h\gamma_{\mathrm{TM}}}{k_h}\right) \end{pmatrix}^{-1}$$
(13)

The ABCD matrix $\mathbf{Q}$ of the WM slab (non-local WM between the interfaces in Fig. 2) as the transmission line supporting only the propagating TEM WM mode with the wavenumber $k_h$ is given by [68],

$$\mathbf{Q} = \begin{pmatrix} \cos(k_h L) & j\eta_h \sin(k_h L) \\ \dfrac{j}{\eta_h}\sin(k_h L) & \cos(k_h L) \end{pmatrix}. \quad (14)$$

Then the global ABCD matrix $\mathbf{M}_g$ for a WM structure shown in Fig. 2 is obtained by cascading the ABCD matrices (12)-(14),

$$\mathbf{M}_g = \mathbf{M}_1 \cdot \mathbf{Q} \cdot \mathbf{M}_2. \quad (15)$$

The global ABCD matrix $\mathbf{M}_g$ for a WM slab without impedance surfaces ($Y_g = 0$ and $\alpha = 0$) is particularly simple,

$$\mathbf{M}_g = \begin{pmatrix} 1 & -\dfrac{j\eta_h \gamma_{\mathrm{TM}}}{k_h}\dfrac{k_x^2}{k_p^2} \\ 0 & \dfrac{k_p^2 + k_x^2}{k_p^2} \end{pmatrix} \cdot \begin{pmatrix} \cos(k_h L) & j\eta_h \sin(k_h L) \\ \dfrac{j}{\eta_h}\sin(k_h L) & \cos(k_h L) \end{pmatrix} \cdot \begin{pmatrix} 1 & \dfrac{j\eta_h \gamma_{\mathrm{TM}}}{k_h}\dfrac{k_x^2}{k_p^2} \\ 0 & \dfrac{k_p^2 + k_x^2}{k_p^2} \end{pmatrix}^{-1}. \quad (16)$$

It should be noted that the proposed ABCD-matrix approach can be used for WM structures terminated with two different impedance surfaces at the interfaces $z = 0$ and $z = L$ corresponding to different conditions on $\alpha$ and different closed-form expressions for $Y_g$.

An interesting observation is that, although we consider reciprocal and lossless media, the ABCD matrices $\mathbf{M}_1$ and $\mathbf{M}_2$ at both interfaces apparently violates reciprocity and seemingly do not obey the conservation of energy (using (10), (13) it can be shown that $\det \mathbf{M}_{1,2} = \dfrac{k_p^2 + k_x^2}{k_p^2}$, rather than unity). Moreover, it can be shown that $\mathrm{Re}\{E_{x,\mathrm{d}} H_{y,\mathrm{d}}^*\} = \dfrac{k_p^2 + k_x^2}{k_p^2}\mathrm{Re}\{E_{x,\mathrm{TEM}} H_{y,\mathrm{TEM}}^*\}$ at the first interface at $z = 0$, which apparently corresponds to the loss in the system, and $\mathrm{Re}\{E_{x,\mathrm{TEM}} H_{y,\mathrm{TEM}}^*\} = \dfrac{k_p^2}{k_p^2 + k_x^2}\mathrm{Re}\{E_{x,\mathrm{d}} H_{y,\mathrm{d}}^*\}$ at the second interface at $z = L$ (Fig. 2), which apparently corresponds to the gain in the system. The same result is easily verified for a WM slab without impedance surface terminations, and it can be shown that in general it does not depend on the termination and it is a property of the wire medium.

To verify the above conclusions we consider the matrix equation (9) with the ABCD matrix $\mathbf{M}_1$ defined by (10). Then, for a lossless reactive impedance surface, $Y_g = j\,\mathrm{Im}\{Y_g\}$ and $\alpha$ real valued,

$$\mathrm{Re}\{E_{x,\mathrm{d}} H_{y,\mathrm{d}}^*\} =$$
$$\mathrm{Re}\left\{ \begin{array}{l} \left[\left(1 + \dfrac{k_x^2}{k_p^2}\dfrac{\alpha\gamma_{\mathrm{TM}}}{(1+\alpha\gamma_{\mathrm{TM}})}\right)\left(1 + \dfrac{k_x^2}{k_p^2}\dfrac{1}{(1+\alpha\gamma_{\mathrm{TM}})}\left(1 - (-j\,\mathrm{Im}\{Y_g\})\dfrac{(-j)\eta_h\gamma_{\mathrm{TM}}}{k_h}\right)\right)\right] \\ \times E_{x,\mathrm{TEM}} H_{y,\mathrm{TEM}}^* \\ + \left((-j\,\mathrm{Im}\{Y_g\}) + (-j)\dfrac{k_h}{\eta_h}\dfrac{k_x^2}{k_p^2}\dfrac{\alpha}{(1+\alpha\gamma_{\mathrm{TM}})}\left(1 - (-j\,\mathrm{Im}\{Y_g\})\dfrac{(-j)\eta_h\gamma_{\mathrm{TM}}}{k_h}\right)\right) \cdot \\ \times \left(-\dfrac{j\eta_h\gamma_{\mathrm{TM}}}{k_h}\dfrac{k_x^2}{k_p^2}\dfrac{1}{(1+\alpha\gamma_{\mathrm{TM}})}\right) E_{x,\mathrm{TEM}}^* H_{y,\mathrm{TEM}} \end{array} \right\}$$
(17)

In (17) the terms $E_{x,\mathrm{TEM}} E_{x,\mathrm{TEM}}^*$ and $H_{y,\mathrm{TEM}} H_{y,\mathrm{TEM}}^*$ are not considered because the coefficients with respect to these terms are imaginary. The result (17) can be simplified as follows:

$$\mathrm{Re}\{E_{x,\mathrm{d}} H_{y,\mathrm{d}}^*\} = \mathrm{Re}\left\{\left(1 + \dfrac{k_x^2}{k_p^2}\dfrac{1}{(1+\alpha\gamma_{\mathrm{TM}})} + \dfrac{k_x^2}{k_p^2}\dfrac{\alpha\gamma_{\mathrm{TM}}}{(1+\alpha\gamma_{\mathrm{TM}})}\right) E_{x,\mathrm{TEM}} H_{y,\mathrm{TEM}}^*\right\}$$
$$+ \mathrm{Re}\left\{ \begin{array}{l} \left(\dfrac{k_x^2}{k_p^2}\dfrac{1}{(1+\alpha\gamma_{\mathrm{TM}})}\mathrm{Im}\{Y_g\}\dfrac{\eta_h\gamma_{\mathrm{TM}}}{k_h} + \alpha\gamma_{\mathrm{TM}}\left(\dfrac{k_x^2}{k_p^2}\dfrac{1}{(1+\alpha\gamma_{\mathrm{TM}})}\right)^2\right) \\ + \alpha\gamma_{\mathrm{TM}}\left(\dfrac{k_x^2}{k_p^2}\dfrac{1}{(1+\alpha\gamma_{\mathrm{TM}})}\right)^2 \mathrm{Im}\{Y_g\}\dfrac{\eta_h\gamma_{\mathrm{TM}}}{k_h} \\ \times \left(E_{x,\mathrm{TEM}} H_{y,\mathrm{TEM}}^* - \left(E_{x,\mathrm{TEM}} H_{y,\mathrm{TEM}}^*\right)^*\right) \end{array} \right\}$$
$$= \dfrac{k_p^2 + k_x^2}{k_p^2}\mathrm{Re}\{E_{x,\mathrm{TEM}} H_{y,\mathrm{TEM}}^*\}. \quad (18)$$

A similar analysis can be shown for the second interface at $z = L$ demonstrating the "apparent" gain in the system. However, the overall response of a bounded WM structure described by the global matrix $\mathbf{M}_g$ (15) (or (16) for a WM slab) is consistent with the lossless property maintaining conservation of energy and reciprocity.

The reason for these apparent violations of conservation of energy is that in the non-local WM the Poynting vector has an additional correction term corresponding to a "hidden power" [57]. Specifically, the Poynting vector associated with the TEM WM mode is *not* given by $1/2\,\mathrm{Re}\{E_{x,\mathrm{TEM}} H_{y,\mathrm{TEM}}^*\}$, but has rather an additional term due to the "additional potential" and current associated with the wire medium structure [57]. As shown in Appendix A, the additional term of the Poynting vector is precisely $\dfrac{k_x^2}{2k_p^2}\mathrm{Re}\{E_{x,\mathrm{TEM}} H_{y,\mathrm{TEM}}^*\}$. Thereby, the relation $\mathrm{Re}\{E_{x,\mathrm{d}} H_{y,\mathrm{d}}^*\} = \dfrac{k_p^2 + k_x^2}{k_p^2}\mathrm{Re}\{E_{x,\mathrm{TEM}} H_{y,\mathrm{TEM}}^*\}$ discussed earlier does not express a violation of the conservation of energy, but rather the continuity of the normal component of the Poynting vector across the interface. Similar arguments can be used to justify the apparent violation of the reciprocity at the WM interface.

The proposed formalism of the equivalent ABCD matrix at the WM interface captures the non-local effects at the interface and in the WM material and correctly models the "hidden power" in the non-local WM.

### B. Equivalent Interface for Two-Sided Non-Local Wire Media

Consider two identical semi-infinite wire media connected through an impedance surface at the interface as shown in Fig. 3. The host permittivity in both WM is $\varepsilon_h$, and it is assumed that the TEM WM mode is the incident field on the either side of the interface.

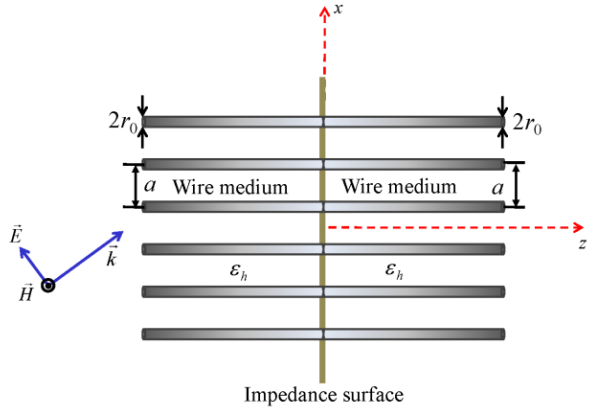

Fig. 3. Two semi-infinite WM connected by an impedance surface at the interface.

At the interface ($z = 0$) the fields satisfy the continuity of the tangential electric-field components, jump condition for the tangential magnetic-field components, and the generalized ABCs for the surface current densities in the WM 1 and 2 [34], [51]:

$$E_{x1,\text{TEM}} + E_{x1,\text{TM}} = E_{x2,\text{TEM}} + E_{x2,\text{TM}} \quad (19a)$$

$$H_{y1,\text{TEM}} + H_{y1,\text{TM}} = H_{y2,\text{TEM}} + H_{y2,\text{TM}} + Y_g(E_{x2,\text{TEM}} + E_{x2,\text{TM}}) \quad (19b)$$

$$\alpha\left(\frac{dJ_{z1,\text{wm}}(z)}{dz} + \frac{dJ_{z2,\text{wm}}(z)}{dz}\right) + J_{z1,\text{wm}}(z) - J_{z2,\text{wm}}(z) = 0 \quad (19c)$$

$$\frac{dJ_{z1,\text{wm}}(z)}{dz} = \frac{dJ_{z2,\text{wm}}(z)}{dz}. \quad (19d)$$

Here, $Y_g$ is the surface admittance of an impedance surface, and the parameter $\alpha$ depends on the material properties of the impedance surface ($\alpha = C_p/2C_w$ for periodic metallic patches [56] and $\alpha = \sigma_s/2j\omega\varepsilon_0\varepsilon_h$ for thin metal/2D material at the interface of two WM [34]). Following the procedure for deriving the ABCD matrix presented in Section II.A, the ABCD matrix for an equivalent interface of two WM connected by an impedance surface can be obtained (with the analytical details provided in appendix B),

$$\begin{pmatrix} E_{x1,\text{TEM}} \\ H_{y1,\text{TEM}} \end{pmatrix} = \mathbf{M} \cdot \begin{pmatrix} E_{x2,\text{TEM}} \\ H_{y2,\text{TEM}} \end{pmatrix} \quad (20)$$

where

$$\mathbf{M} = \begin{pmatrix} 1 & 0 \\ m_{21} & 1 \end{pmatrix} \quad (21)$$

and

$$m_{21} = \frac{Y_g + j\omega\varepsilon_0\varepsilon_h \alpha \dfrac{k_x^2}{k_p^2(1+\alpha\gamma_{\text{TM}})}\left(2 - \dfrac{j\gamma_{\text{TM}}Y_g}{\omega\varepsilon_0\varepsilon_h}\right)}{1 + \dfrac{k_x^2}{2k_p^2(1+\alpha\gamma_{\text{TM}})}\left(2 - \dfrac{j\gamma_{\text{TM}}Y_g}{\omega\varepsilon_0\varepsilon_h}\right)}. \quad (22)$$

It can be seen that in the limiting case of $Y_g = 0$ and $\alpha = 0$, $m_{21} = 0$.

The resulting ABCD matrix (21) is consistent with the usual reciprocity constraint (det $\mathbf{M} = 1$), and for a lossless reactive impedance surface, $Y_g = j\text{Im}\{Y_g\}$ and $\alpha$ real-valued, such that $m_{21} = j\text{Im}\{m_{21}\}$ it also satisfies the usual condition of conservation of energy,

$$\begin{aligned}
\text{Re}\{E_{x1,\text{TEM}}H_{y1,\text{TEM}}^*\} &= \text{Re}\{E_{x2,\text{TEM}}(m_{21}^*E_{x2,\text{TEM}}^* + H_{y2,\text{TEM}}^*)\} \\
&= \text{Re}\{E_{x2,\text{TEM}}H_{y2,\text{TEM}}^*\}.
\end{aligned} \quad (23)$$

It should be noted that both WM 1 and WM 2 (Fig. 3) have an additional correction term in the Poynting vector representation as it has been proven in [57]. However, because in the case presented here the wire media 1 and 2 are identical, the additional Poynting vector terms turn out to be also identical, and thereby cancel out. For two different wire media (different geometrical and host material parameters) connected at the interface by an impedance surface the ABCD matrix will apparently violate reciprocity and conservation of energy. This case is omitted here because the generalized two-sided ABCs (19c), (19d) have not been derived for a general case of two wire media connected to an arbitrary impedance surface and having different lattice periods, radii of the wires, and host permittivities. The only case of generalized two-sided ABCs has been considered in [34] with two wire media having different host permittivities connected by a thin metal/2D material at the interface. Following the formulation presented here to derive the ABCD matrix with the generalized ABCs in [34] (with two different host permittivities) it can be shown that the ABCD matrix will be seemingly non-reciprocal and violate the conservation of energy. However, in all bounded WM structures there are at least two interfaces, such that if one of them provides apparent loss (as the wave enters the WM), the other interface provides apparent gain (as the wave exits the WM), so that the overall response is consistent with the lossless property maintaining conservation of energy and reciprocity.

Also, a special case of interest here is a continuous impedance surface at the interface, $Y_g = \sigma_s$, where in general the surface conductivity $\sigma_s$ can be complex-valued. Then with $\alpha = \sigma_s/2j\omega\varepsilon_0\varepsilon_h$ it can be shown that (22) reduces to $m_{21} = \sigma_s$. In this case there is no TM WM mode contribution in the ABCD matrix for the equivalent interface and the problem is completely described by the TEM WM mode only.

With the formalism of an equivalent transmission-network approach presented here for two cases of WM interfacing a local dielectric material and two WM connected by an impedance surface enables to model various geometrically-complex multilayer WM structures (with in general different impedance surfaces at the interfaces) by cascading the obtained above ABCD matrices of equivalent interfaces and ABCD matrices of WM slabs as transmission lines supporting the only propagating TEM WM mode.

In the section to follow several representative numerical examples will be given based on the proposed equivalent transmission-network formulation with the results compared to the non-local solution.

## III. NUMERICAL RESULTS AND DISCUSSIONS

The numerical analysis is carried out based on the proposed equivalent transmission-network approach and the results are compared to the non-local solution for several representative examples. The non-local homogenization model have been extensively verified with full-wave numerical simulations for various WM topologies [5]-[13], [15]-[18], [29], [30], [46]-[53], and can be used for an adequate comparison with the ABCD-matrix approach. In all the examples the obliquely incident TM-polarized plane wave is considered for excitation, where the reflection and transmission coefficients (S-matrix) are retrieved from the ABCD-matrix parameters with the known expressions from the microwave engineering [68, p. 192; with $Z_0 = \eta_h$].

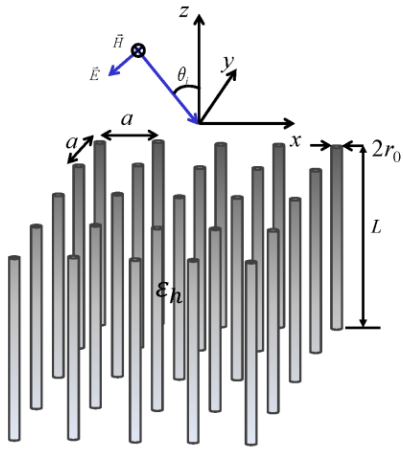

Fig. 4. Geometry of a WM slab in air with the obliquely incident TM-polarized plane wave.

In the first example, a WM slab in air is considered (with the geometry shown in Fig. 4). The response of the structure (reflection and transmission) is studied for the following geometrical and material parameters of WM: $k_0 a = 1$, $r_0/a = 0.05$, $\varepsilon_h = 2$, $\theta_i = 75^o$. The global ABCD matrix (16) is used for the calculation of the reflection coefficient $S_{11}$ and transmission coefficient $S_{21}$ at the WM interfaces, respectively, with the results for the magnitudes shown in Fig. 5 and compared with those based on the non-local solution. The non-local homogenization model for a WM slab has been extensively presented in the literature (see for example [15] with the analytical details for reflection and transmission coefficients). The results are shown versus $L/a$ demonstrating nearly perfect agreement with the non-local results for $L/a > 2$, when the WM interfaces are decoupled by the evanescent TM WM mode. In other words, the developed ABCD formalism yields rather accurate results when it is possible to neglect the effects of the "incident" evanescent TM waves inside the WM. Note that the non-local effects due to the TM WM mode excited by the TEM waves propagating inside the WM are captured by the transmission matrices for equivalent WM interfaces.

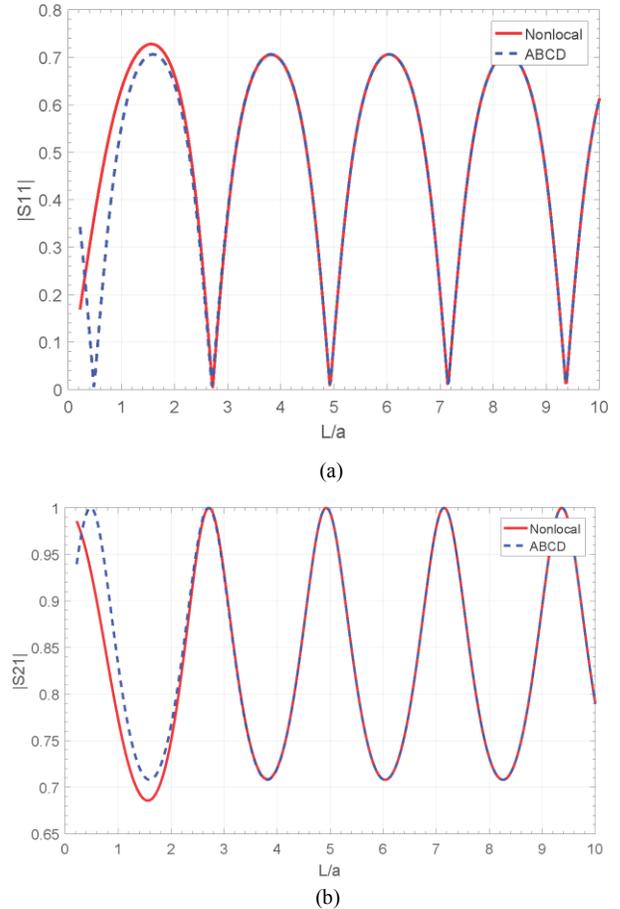

Fig. 5. Magnitude of the (a) reflection coefficient $|S_{11}|$ and (b) transmission coefficient $|S_{21}|$ versus $L/a$ for a WM slab in air. The ABCD-matrix results are compared with the non-local solution showing excellent agreement for $L/a > 2$.

In the second example, a two-sided mushroom topology is considered with the wires terminated with patch arrays at the WM interfaces (with the geometry shown in Fig. 6).

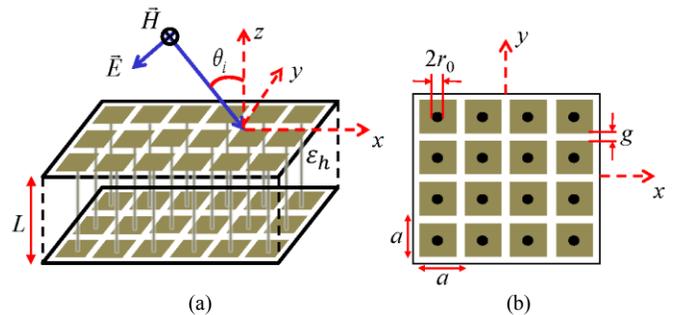

Fig. 6. (a) Geometry of a two-sided mushroom structure in air with the patch arrays at the WM interfaces excited by an obliquely incident TM-polarized plane wave and (b) top view of the patch array connected to wires.

The following geometrical and material parameters are used in the calculations: $k_0 a = 1$, $r_0/a = 0.05$, $g/a = 0.1$, $\varepsilon_h = 2$, $\theta_i = 75^o$. The global ABCD matrix (15) (with the ABCD matrices for the equivalent interfaces (12) and (13) and the ABCD matrix of the WM slab (14) supporting the TEM WM mode) is used to determine the reflection and transmission coefficients. The results for $|S_{11}|$ and $|S_{21}|$ versus $L/a$ are shown in Fig. 7. It can be seen that the agreement between the ABCD results and non-local results (with the analytical details for a non-local model given in [64]) even better than for a WM slab (previous example) for a smaller ratio $L/a$ due to a stronger confinement of the evanescent TM WM mode at the interface because of the presence of the patch array.

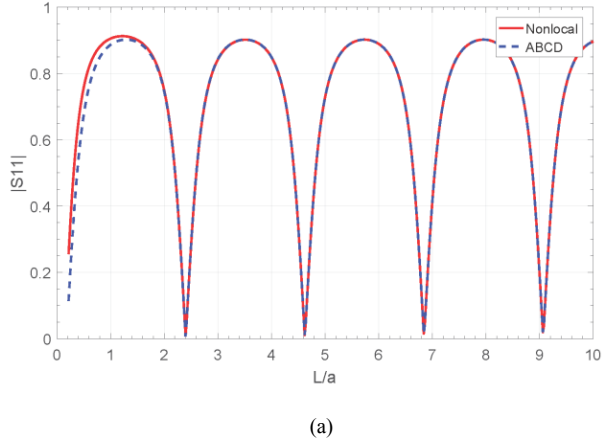

(a)

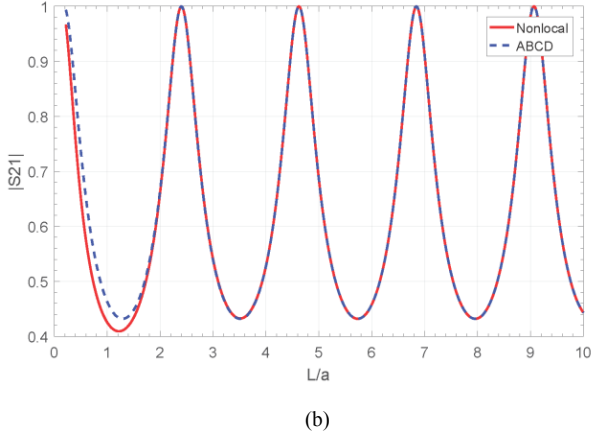

(b)

Fig. 7. Magnitude of the (a) reflection coefficient $|S_{11}|$ and (b) transmission coefficient $|S_{21}|$ versus $L/a$ for a two-sided mushroom structure. The ABCD-matrix results are compared with the non-local solution showing excellent agreement for $L/a > 1.8$.

In the final example (with the geometry shown in Fig. 8) we consider a multilayer mushroom topology comprised of 4 WM slabs and 5 patch arrays at the WM interfaces. An obliquely incident TM-polarized plane at 75 degrees from the air region is used for the excitation. Each WM slab is air-filled with $\varepsilon_h = 1$, having thickness $L = 2$ mm with the period of wires (and the patches) $a = 1$ mm, radius of wires $r_0 = 0.05$ mm, and gap between the patches $g = 0.1$ mm. The global ABCD matrix is obtained by cascading the ABCD matrices of the top equivalent interface (12), WM slabs (14), two-sided equivalent interfaces (21), (22), and the bottom equivalent interface (13). Then the global ABCD matrix is used to calculate the S-parameters at the top and bottom interfaces. The results for the magnitudes of the reflection coefficient $|S_{11}|$ and transmission coefficient $|S_{21}|$ versus frequency are shown in Fig. 9 and compared with the non-local results (with the non-local formulation given in [6], [7]) demonstrating perfect agreement in the entire frequency range.

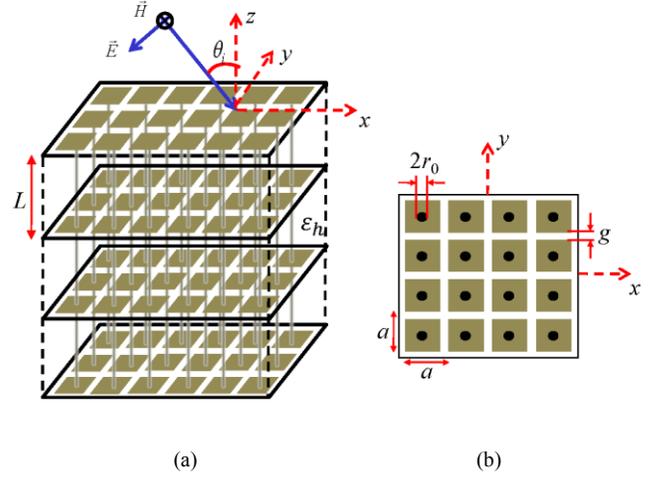

(a) (b)

Fig. 8. (a) Geometry of a multilayered mushroom structure with an obliquely incident TM-polarized plane wave and (b) top view of the patch array connected to wires.

Also, we have considered an infinite periodic structure comprised of WM slabs with the patch arrays at the interfaces, and studied Bloch waves propagating along the $z$-direction. The dispersion equation is obtained as follows:

$$AD + e^{2\gamma_b L} - (A+D)e^{\gamma_b L} - BC = 0 \qquad (24)$$

where $\gamma_b = \alpha_b + j\beta_b$ is the propagation constant of Bloch waves of the periodic WM structure, and the ABCD parameters are obtained by cascading the ABCD matrices of WM slabs (of thickness $L/2$) and the equivalent interface for a two-sided WM connected by an impedance surface (21), (22),

$$\begin{bmatrix} A & B \\ C & D \end{bmatrix} =$$
$$\begin{bmatrix} \cos\left(k_h \frac{L}{2}\right) & j\eta_h \sin\left(k_h \frac{L}{2}\right) \\ \frac{j}{\eta_h}\sin\left(k_h \frac{L}{2}\right) & \cos\left(k_h \frac{L}{2}\right) \end{bmatrix} \cdot \begin{bmatrix} 1 & 0 \\ m_{21} & 1 \end{bmatrix} \cdot \begin{bmatrix} \cos\left(k_h \frac{L}{2}\right) & j\eta_h \sin\left(k_h \frac{L}{2}\right) \\ \frac{j}{\eta_h}\sin\left(k_h \frac{L}{2}\right) & \cos\left(k_h \frac{L}{2}\right) \end{bmatrix}.$$

(25)

The solution of (24) with the reciprocity condition for the ABCD matrix (25) ($AD - BC = 1$) results in closed-form expression for the phase constant of Bloch waves,

$$\beta_b L = \text{Im}\left\{\cosh^{-1}\left(\frac{A+D}{2}\right)\right\}. \qquad (26)$$

The numerical results of (26) are superimposed in Fig. 9(b) as Brillouin diagram. It can be seen that transmission resonances of the finite multilayer structure correspond to the passbands of the infinite structure, and the rejection band in the finite structure is well approximated by the stopband of an infinite structure. This observation is consistent with the results obtained previously for stacked periodic 2D metallic meshes [69] and periodic 2D conducting patches [70] at microwave frequencies, and in a graphene-dielectric microstructure at low-terahertz frequencies [71].

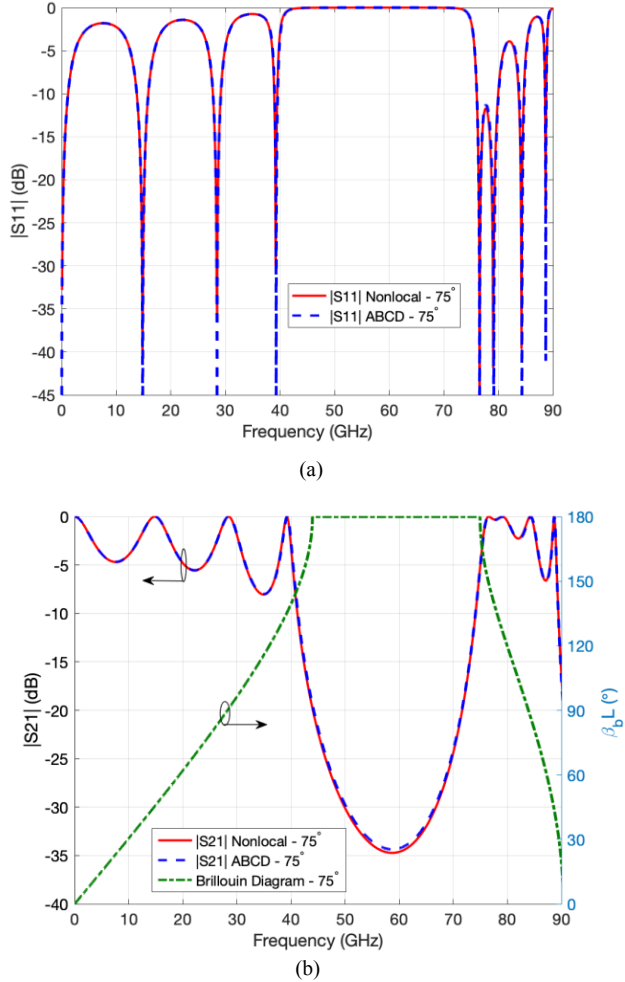

Fig. 9. Magnitude of the (a) reflection coefficient $|S_{11}|$ and (b) transmission coefficient $|S_{21}|$ versus frequency for a multilayer mushroom structure. Brillouin diagram is also depicted for Bloch waves of an infinite periodic structure of WM slabs connected to patch arrays at the interfaces.

## IV. CONCLUSION

We proposed a circuit-model formalism for non-local bounded WM structures with arbitrary terminations. We introduced the idea of an equivalent interface and derived the transmission network for a semi-infinite WM interfacing a local dielectric material, and then generalized the formalism for two non-local WM connected by an impedance surface. It is observed that the ABCD matrix for the equivalent interface apparently violates the conservation of energy and reciprocity, and seemingly behaves as a non-reciprocal lossy or active system. However, for bounded WM structure having at least two interfaces the overall response is consistent with the lossless property maintaining conservation of energy and reciprocity. We demonstrated that these exotic features are due to the non-standard expression of the Poynting vector in the non-local material. We have demonstrated that in the non-local WM the Poynting vector has an additional correction term corresponding to "hidden power" due to non-local effects, and the results obtained here for the additional term are consistent with previously published results.

The equivalent transmission-network formalism presented in the paper enables to model multilayer WM configurations with an arbitrary number of WM layers terminated with in general arbitrary impedance surfaces. For such a general case there is no analytical model available in the literature, and the approach presented in the paper is the first attempt to solve this problem.

The results presented in the paper have been verified with the non-local solution for several representative examples showing nearly perfect agreement subject to the condition that WM interfaces are decoupled by the evanescent TM WM mode below the plasma frequency. Indeed, the model yields nearly exact results when the WM interfaces are sufficiently far apart so that the "incident" evanescent TM WM modes have a negligible effect.

## APPENDIX A

ADDITIONAL CORRECTION TERM IN THE POYNTING VECTOR FOR WM

According to [57, eq. (66)], the time-averaged Poynting vector for the case of fields with a spatial dependence of the form $e^{-j\mathbf{k}\cdot\mathbf{r}}$ with $\mathbf{k}$ real-valued and for a lossless spatially dispersive WM in the $z$-direction is given by the following expression,

$$\mathbf{S}_{ave,z} = \frac{1}{2}\text{Re}\left\{\left(\mathbf{E}_{wm}\times\mathbf{H}_{wm}^*\right)_z\right\} - \frac{\omega}{4}\mathbf{E}_{wm}^* \cdot \frac{\partial \overset{\text{t}}{\varepsilon}(\omega,\mathbf{k})}{\partial k_z} \cdot \mathbf{E}_{wm} \quad (27)$$

where the effective dielectric function is defined as

$$\overset{\text{t}}{\varepsilon}(\omega,\mathbf{k}) = \varepsilon_0 \varepsilon_h \left(\overset{\text{t}}{I} - \hat{\mathbf{z}}\hat{\mathbf{z}} \frac{k_p^2}{k_h^2 - k_z^2}\right) \quad (28)$$

and $k_z$ is the $z$-component of the wave vector $\mathbf{k}=(k_x,0,k_z)$.

The additional term in (27) can be written as follows:

$$\frac{\omega}{4}\mathbf{E}_{wm}^* \cdot \frac{\partial \overset{\text{t}}{\varepsilon}(\omega,\mathbf{k})}{\partial k_z} \cdot \mathbf{E}_{wm} = \frac{\omega}{4}E_{z,\text{TEM}}^* \frac{\partial \varepsilon_{zz}(\omega,\mathbf{k})}{\partial k_z} E_{z,\text{TEM}} \quad (29)$$

and using $\nabla\cdot\mathbf{D}_{wm}=\nabla\cdot(\overset{\text{t}}{\varepsilon}\cdot\mathbf{E}_{wm})=0$ such that $\frac{E_{z,\text{TEM}}}{k_h^2-k_z^2}=\frac{k_x}{k_p^2 k_h}E_{x,\text{TEM}}$, and with the derivative of (28), $\frac{\partial \varepsilon_{zz}(\omega,\mathbf{k})}{\partial k_z}=\frac{\varepsilon_0 \varepsilon_h k_p^2(-2k_z)}{\left(k_h^2-k_z^2\right)^2}$ results in

$$\frac{\omega}{4}E_{z,\text{TEM}}^*\frac{\partial \varepsilon_{zz}(\omega,\mathbf{k})}{\partial k_z}E_{z,\text{TEM}}=\frac{\omega}{4}\frac{k_x}{k_p^2 k_h}E_{x,\text{TEM}}^*(-2\varepsilon_0\varepsilon_h k_p^2 k_h)\frac{k_x}{k_p^2 k_h}E_{x,\text{TEM}}. \quad (30)$$

By relating $\frac{E_{x,\text{TEM}}}{H_{y,\text{TEM}}}=Z_{\text{TEM}}=\frac{k_h}{\omega\varepsilon_0\varepsilon_h}$ we obtain

$$\frac{\omega}{4}E_{z,\text{TEM}}^*\frac{\partial \varepsilon_{zz}(\omega,\mathbf{k})}{\partial k_z}E_{z,\text{TEM}}=-\frac{1}{2}\frac{k_x^2}{k_p^2}E_{x,\text{TEM}}H_{y,\text{TEM}}^*. \quad (31)$$

Also, by considering the additional term in [57, eq. (65)] with the Poynting vector in the $z$-direction and assuming only the TEM WM mode results in

$$\frac{\varphi_w I_{z,\text{wm}}^*}{A_c}=-\frac{A_c}{j\omega C_w}\frac{\partial J_{z,\text{wm}}}{\partial z}J_{z,\text{wm}}^* \quad (32)$$

where $J_{z,\text{wm}}^*=\frac{I_{z,\text{wm}}^*}{A_c}$ and $\frac{\partial J_{z,\text{wm}}}{\partial z}=-\frac{j\omega C_w}{A_c}\varphi_w$, and in terms of field components, $J_{z,\text{wm}}^* = jk_x H_{y,\text{TEM}}^*$ and $\dfrac{\partial J_{z,\text{wm}}}{\partial z} = -k_x \omega \varepsilon_0 \varepsilon_h E_{x,\text{TEM}}$.

Then,

$$\frac{\varphi_w I_{z,\text{wm}}^*}{A_c} = -\frac{A_c}{j\omega C_w}(-k_x \omega \varepsilon_0 \varepsilon_h) E_{x,\text{TEM}}(jk_x)H_{y,\text{TEM}}^* = \frac{k_x^2}{k_p^2} E_{x,\text{TEM}} H_{y,\text{TEM}}^* \quad (33)$$

where $k_p^2 = \dfrac{C_w}{A_c \varepsilon_0 \varepsilon_h}$.

The results (31) and (33) for the additional term in the Poynting vector expressions (66) and (65) in Ref. [57] are consistent with the power representation based on the equivalent transmission-network analysis considered in this paper.

## APPENDIX B

DERIVATION OF THE ABCD MATRIX FOR AN EQUIVALENT INTERFACE OF TWO WM CONNECTED BY AN IMPEDANCE SURFACE

The ABC (19c) can be expressed in terms of the field components in the WM 1 and 2 as follows:

$$k_x H_{y1,\text{wm}} + \omega \varepsilon_0 \varepsilon_h E_{z1,\text{wm}} + \alpha\left(k_x \frac{\partial H_{y1,\text{wm}}}{\partial z} + \omega \varepsilon_0 \varepsilon_h \frac{\partial E_{z1,\text{wm}}}{\partial z}\right) \\ = k_x H_{y2,\text{wm}} + \omega \varepsilon_0 \varepsilon_h E_{z2,\text{wm}} - \alpha\left(k_x \frac{\partial H_{y2,\text{wm}}}{\partial z} + \omega \varepsilon_0 \varepsilon_h \frac{\partial E_{z2,\text{wm}}}{\partial z}\right). \quad (34)$$

Taking into account (3) results in

$$E_{z1,2,\text{TM}} = -\frac{1}{\omega \varepsilon_0 \varepsilon_h} \frac{k_p^2 + k_x^2}{k_x} H_{y1,2,\text{TM}}. \quad (35)$$

Then, from Maxwell's equations and with the assumption that there is no incident TM WM mode from either side on the interface at $z = 0$ (Fig. 3),

$$E_{x1,\text{TM}} = -\frac{1}{j\omega \varepsilon_0 \varepsilon_h} \frac{\partial H_{y1,\text{TM}}}{\partial z} = \frac{j\gamma_{\text{TM}}}{\omega \varepsilon_0 \varepsilon_h} H_{y1,\text{TM}} \\ E_{x2,\text{TM}} = -\frac{1}{j\omega \varepsilon_0 \varepsilon_h} \frac{\partial H_{y2,\text{TM}}}{\partial z} = -\frac{j\gamma_{\text{TM}}}{\omega \varepsilon_0 \varepsilon_h} H_{y2,\text{TM}} \quad (36)$$

and that $\dfrac{\partial E_{z1,\text{wm}}}{\partial z} \equiv \dfrac{\partial E_{z1,\text{TM}}}{\partial z} = -\dfrac{\gamma_{\text{TM}}}{\omega \varepsilon_0 \varepsilon_h} \dfrac{k_p^2 + k_x^2}{k_p^2} H_{y1,\text{TM}}$,

$\dfrac{\partial E_{z2,\text{wm}}}{\partial z} \equiv \dfrac{\partial E_{z2,\text{TM}}}{\partial z} = \dfrac{\gamma_{\text{TM}}}{\omega \varepsilon_0 \varepsilon_h} \dfrac{k_p^2 + k_x^2}{k_p^2} H_{y2,\text{TM}}$, the ABC (34) can be written as follows:

$$k_x(H_{y1,\text{TEM}} + H_{y1,\text{TM}}) - \frac{k_p^2 + k_x^2}{k_x} H_{y1,\text{TM}} \\ + \alpha\left(-j\omega \varepsilon_0 \varepsilon_h k_x(E_{x1,\text{TEM}} + E_{x1,\text{TM}}) - \gamma_{\text{TM}} \frac{k_p^2 + k_x^2}{k_x} H_{y1,\text{TM}}\right) \\ = k_x(H_{y2,\text{TEM}} + H_{y2,\text{TM}}) - \frac{k_p^2 + k_x^2}{k_x} H_{y2,\text{TM}} \\ - \alpha\left(-j\omega \varepsilon_0 \varepsilon_h k_x(E_{x2,\text{TEM}} + E_{x2,\text{TM}}) + \gamma_{\text{TM}} \frac{k_p^2 + k_x^2}{k_x} H_{y2,\text{TM}}\right). \quad (37)$$

Substituting (36) in (37) and after simplifications we obtain,

$$k_x^2 H_{y1,\text{TEM}} - k_p^2(1 + \alpha\gamma_{\text{TM}})H_{y1,\text{TM}} - j\omega \varepsilon_0 \varepsilon_h \alpha k_x^2 E_{x1,\text{TEM}} \\ = k_x^2 H_{y2,\text{TEM}} - k_p^2(1 + \alpha\gamma_{\text{TM}})H_{y2,\text{TM}} + j\omega \varepsilon_0 \varepsilon_h \alpha k_x^2 E_{x2,\text{TEM}}. \quad (38)$$

The ABC (19d) can be expressed in terms of the field components in WM 1 and 2,

$$k_x \frac{\partial H_{y1,\text{wm}}}{\partial z} + \omega \varepsilon_0 \varepsilon_h \frac{\partial E_{z1,\text{wm}}}{\partial z} = k_x \frac{\partial H_{y2,\text{wm}}}{\partial z} + \omega \varepsilon_0 \varepsilon_h \frac{\partial E_{z2,\text{wm}}}{\partial z} \quad (39)$$

and taking into account (35) and (36) the ABC (39) can be written as

$$-j\omega \varepsilon_0 \varepsilon_h k_x\left(E_{x1,\text{TEM}} + \frac{j\gamma_{\text{TM}}}{\omega \varepsilon_0 \varepsilon_h} H_{y1,\text{TM}}\right) - \gamma_{\text{TM}} \frac{k_p^2 + k_x^2}{k_x} H_{y1,\text{TM}} \\ = -j\omega \varepsilon_0 \varepsilon_h k_x\left(E_{x2,\text{TEM}} - \frac{j\gamma_{\text{TM}}}{\omega \varepsilon_0 \varepsilon_h} H_{y2,\text{TM}}\right) + \gamma_{\text{TM}} \frac{k_p^2 + k_x^2}{k_x} H_{y2,\text{TM}} \quad (40)$$

which results in

$$-j\omega \varepsilon_0 \varepsilon_h k_x^2 E_{x1,\text{TEM}} - \gamma_{\text{TM}} k_p^2 H_{y1,\text{TM}} = -j\omega \varepsilon_0 \varepsilon_h k_x^2 E_{x2,\text{TEM}} + \gamma_{\text{TM}} k_p^2 H_{y2,\text{TM}}. \quad (41)$$

By solving the system of equations (38) and (41) for $H_{y1,\text{TM}}$ and $H_{y2,\text{TM}}$ we obtain,

$$H_{y1,\text{TM}} = \frac{k_x^2}{2k_p^2(1 + \alpha\gamma_{\text{TM}})}\left(H_{y1,\text{TEM}} - H_{y2,\text{TEM}}\right) \\ - \frac{j\omega \varepsilon_0 \varepsilon_h}{\gamma_{\text{TM}}} \frac{k_x^2}{2k_p^2(1 + \alpha\gamma_{\text{TM}})}\left(E_{x1,\text{TEM}}(1 + 2\alpha\gamma_{\text{TM}}) - E_{x2,\text{TEM}}\right) \quad (42)$$

$$H_{y2,\text{TM}} = -\frac{k_x^2}{2k_p^2(1 + \alpha\gamma_{\text{TM}})}\left(H_{y1,\text{TEM}} - H_{y2,\text{TEM}}\right) \\ - \frac{j\omega \varepsilon_0 \varepsilon_h}{\gamma_{\text{TM}}} \frac{k_x^2}{2k_p^2(1 + \alpha\gamma_{\text{TM}})}\left(E_{x1,\text{TEM}} - (1 + 2\alpha\gamma_{\text{TM}})E_{x2,\text{TEM}}\right). \quad (43)$$

The continuity condition (19a) for tangential electric-field components with the relation (36) can be written as

$$E_{x1,\text{TEM}} + \frac{j\gamma_{\text{TM}}}{\omega \varepsilon_0 \varepsilon_h} H_{y1,\text{TM}} = E_{x2,\text{TEM}} - \frac{j\gamma_{\text{TM}}}{\omega \varepsilon_0 \varepsilon_h} H_{y2,\text{TM}}. \quad (44)$$

Substituting (42), (43) in (44) we obtain,

$$E_{x1,\text{TEM}} + \frac{j\gamma_{\text{TM}}}{\omega \varepsilon_0 \varepsilon_h} \frac{k_x^2}{2k_p^2(1 + \alpha\gamma_{\text{TM}})}\left(H_{y1,\text{TEM}} - H_{y2,\text{TEM}}\right) \\ - \frac{j\gamma_{\text{TM}}}{\omega \varepsilon_0 \varepsilon_h} \frac{j\omega \varepsilon_0 \varepsilon_h}{\gamma_{\text{TM}}} \frac{k_x^2}{2k_p^2(1 + \alpha\gamma_{\text{TM}})}\left(E_{x1,\text{TEM}}(1 + 2\alpha\gamma_{\text{TM}}) - E_{x2,\text{TEM}}\right) \\ = E_{x2,\text{TEM}} + \frac{j\gamma_{\text{TM}}}{\omega \varepsilon_0 \varepsilon_h} \frac{k_x^2}{2k_p^2(1 + \alpha\gamma_{\text{TM}})}\left(H_{y1,\text{TEM}} - H_{y2,\text{TEM}}\right) \\ + \frac{j\gamma_{\text{TM}}}{\omega \varepsilon_0 \varepsilon_h} \frac{j\omega \varepsilon_0 \varepsilon_h}{\gamma_{\text{TM}}} \frac{k_x^2}{2k_p^2(1 + \alpha\gamma_{\text{TM}})}\left(E_{x1,\text{TEM}} - (1 + 2\alpha\gamma_{\text{TM}})E_{x2,\text{TEM}}\right) \quad (45)$$

which after simplifications results in

$$E_{x1,\text{TEM}} = E_{x2,\text{TEM}}. \quad (46)$$

This is an interesting result such that the continuity of WM tangential electric-field components (TEM + TM) at the interface reduces to the continuity of the TEM WM modes only.

Next, substituting (42), (43) in the jump condition (19b) for tangential magnetic-field components gives the following result:

$$H_{y1,\text{TEM}} + \frac{k_x^2}{2k_p^2(1+\alpha\gamma_{\text{TM}})}\left(H_{y1,\text{TEM}} - H_{y2,\text{TEM}}\right)$$
$$- \frac{j\omega\varepsilon_0\varepsilon_h}{\gamma_{\text{TM}}} \frac{k_x^2}{2k_p^2(1+\alpha\gamma_{\text{TM}})}\left(E_{x1,\text{TEM}}(1+2\alpha\gamma_{\text{TM}}) - E_{x2,\text{TEM}}\right)$$
$$= H_{y2,\text{TEM}} - \frac{k_x^2}{2k_p^2(1+\alpha\gamma_{\text{TM}})}\left(H_{y1,\text{TEM}} - H_{y2,\text{TEM}}\right)$$
$$- \frac{j\omega\varepsilon_0\varepsilon_h}{\gamma_{\text{TM}}} \frac{k_x^2}{2k_p^2(1+\alpha\gamma_{\text{TM}})}\left(E_{x1,\text{TEM}} - (1+2\alpha\gamma_{\text{TM}})E_{x2,\text{TEM}}\right)$$
$$+ Y_g E_{x2,\text{TEM}} + \frac{j\gamma_{\text{TM}}Y_g}{\omega\varepsilon_0\varepsilon_h} \frac{k_x^2}{2k_p^2(1+\alpha\gamma_{\text{TM}})}\left(H_{y1,\text{TEM}} - H_{y2,\text{TEM}}\right)$$
$$+ \frac{j\gamma_{\text{TM}}Y_g}{\omega\varepsilon_0\varepsilon_h} \frac{j\omega\varepsilon_0\varepsilon_h}{\gamma_{\text{TM}}} \frac{k_x^2}{2k_p^2(1+\alpha\gamma_{\text{TM}})}\left(E_{x1,\text{TEM}} - (1+2\alpha\gamma_{\text{TM}})E_{x2,\text{TEM}}\right)$$
(47)

which can be simplified further,

$$H_{y1,\text{TEM}}\left(1 + \frac{k_x^2}{2k_p^2(1+\alpha\gamma_{\text{TM}})}\left(2 - \frac{j\gamma_{\text{TM}}Y_g}{\omega\varepsilon_0\varepsilon_h}\right)\right)$$
$$+ E_{x1,\text{TEM}} \frac{j\omega\varepsilon_0\varepsilon_h}{\gamma_{\text{TM}}} \frac{k_x^2}{2k_p^2(1+\alpha\gamma_{\text{TM}})}\left(-2\alpha\gamma_{\text{TM}} - \frac{j\gamma_{\text{TM}}Y_g}{\omega\varepsilon_0\varepsilon_h}\right)$$
$$= H_{y2,\text{TEM}}\left(1 + \frac{k_x^2}{2k_p^2(1+\alpha\gamma_{\text{TM}})}\left(2 - \frac{j\gamma_{\text{TM}}Y_g}{\omega\varepsilon_0\varepsilon_h}\right)\right)$$
$$+ E_{x2,\text{TEM}}\left(Y_g + \frac{j\omega\varepsilon_0\varepsilon_h}{\gamma_{\text{TM}}} \frac{k_x^2}{2k_p^2(1+\alpha\gamma_{\text{TM}})}\left(2\alpha\gamma_{\text{TM}} - \frac{j\gamma_{\text{TM}}Y_g}{\omega\varepsilon_0\varepsilon_h}(1+2\alpha\gamma_{\text{TM}})\right)\right)$$
(48)

The system of equations (46), (48) results in the ABCD matrix (21), (22).